\documentclass{article}
\usepackage{amsmath,amsfonts,color}
\def\R{\mathbb{R}}
\def\C{\mathbb{C}}

\def\pas{\par\smallskip}
\def\pam{\par\medskip}
\def\pamn{\pam\noindent}

\def\ba{\begin{array}}
\def\ea{\end{array}}
\def\mm{\medskip\\}
\def\aru#1{\left\{\ba{l}#1\ea\right.}
\def\zx#1{\begin{equation}\label{#1}}
\def\zc{\end{equation}}

\def\vv#1{{\bf #1}}

\renewcommand{\theequation}{\thesection.\theequation}
\numberwithin{equation}{section}
 \title{Spinors, matrix structures, and projective geometry \\in polarization optics}

    \author{Elena Ovsiyuk\footnote{Mosyr State Pedagogical University,
    Belarus, e.ovsiyuk@mail.ru},
 Olga Veko\footnote{Kalinkovichi Gymnasium, Belarus,vekoolga@mail.ru},
             Mircea Neagu\footnote{Transilvania University of
            Brasov, mircea.neagu@unitbv.ro},\\
            Vladimir Balan\footnote{University Politehnica of Bucharest, Romania, vladimir.balan@upb.ro},
             Victor Red'kov\footnote{B.I. Stepanov Institute
of Physics, NAS of Belarus, redkov@dragon.bas-net.by}}
    \date{}

 \date{}
\begin{document}\maketitle
\begin{abstract}
The paper discusses the role played by Mueller and Jones formalisms in polarization
    optics, by addressing the following aspects: restriction to the $SU(2)$ symmetry,
    non-relativistic Stokes 3-vectors; Cartan 2-spinors in polarization optics;
    Jones 4-spinors for partially polarized light; the linear group $SL(4,\R)$ and
    the classification of 1-parametric Mueller matrices; semi-group structure and
    classification of degenerate Mueller matrices.

\end{abstract}\pas\noindent
{\bf MSC2010}: 15A66, 78A25, 35Q60, 78A99, 81V10.\pas\noindent
{\bf Key-words}: spinors; Mueller formalism; Jones formalism; Stokes 3-vectors;
    Cartan 2-spinors; polarization optics; symmetry.

%
%
\section*{General introduction}
The goal of the paper is to discuss the role played by Mueller (matrix) and
 Jones (spinor) formalisms in polarization optics, by addressing the following
 essential aspects:
\begin{itemize}\setlength{\parskip}{-1mm}
\item polarization of the light and Mueller formalism;
\item polarized light and Jones formalism, restriction to the $SU(2)$ symmetry,
    and two types of non-relativistic Stokes 3-vectors;
\item Cartan 2-spinors in polarization optics: two kinds of Jones complex 2-vectors;
\item on possible Jones 4-spinors for partially polarized light;
\item the linear group $SL(4,\R)$ and the classification of 1-parametric Mueller matrices;
\item  classification of degenerate Mueller matrices with semi-group structure,
    and associated projective transformations.
\end{itemize}
%
%
\section{Polarization of the light and the Mueller formalism}
To elucidate in which way mathematical theory of rotation and Lorentz groups  \cite{1} may be
    applied to problems of polarization optics \cite{3}, and also which problems from this field
    await to be solved, we proceed with basic definitions concerning the light polarization.\pas
Consider a plane electromagnetic wave spreading along the axis $z$; then, at an arbitrary
    fixed point $z$, we have
    $$E^1=N \cos \omega t\; , \qquad E^2=M\cos (\omega t+\Delta)\; , \qquad E^3=0\; ,
    N \geq 0\;, \qquad M \geq 0\; , \qquad \Delta \in [- \pi ,+\pi ]\;,$$
and the Stokes parameters $(S_{a}) =(I, S^1, S^2,S^3)$ are determined by
    $$I =\;<E_1^2+E_2^2 >\; , \qquad S^3=\;< E_1^2- E_2^2 >\; ,
    S^1=\;< 2E_1E_2\; \cos \Delta >\; , \qquad S^2=\;<2E_1E_2\; \sin \Delta >\; ;$$
where $M(t),N(t)$ are amplitudes of two electric components, $\Delta(t)$ is a phase shift and
    the symbol $ <\ldots>$ stands for averaging in time.\pas
If the amplitudes $N(t),M(t)$ and the phase shift $\Delta(t)$ do not substantially depend on time
    (or at all, as in the case of completely polarized light), during the measuring process the
    Stokes parameters equal to
    $$S^{0} _{pol}=I_{pol}=N^2+M^2\; , \qquad S^3_{pol} =N^2-M^2\; ,
    S^1_{pol}=2NM\;\cos \Delta\; , \qquad S^2 _{pol}= 2NM\; \sin \Delta\; ,$$
and the following identity holds
    $$S_{a} S^{a}=I^2_{pol}-{\vec S}^{\;2}_{pol}=0,$$
that is, ${\vec S}=I_{pol}\; {\vec n}$. In other words, for completely polarized light, the
    Stokes 4-vector is isotropic. For the natural (non-polarized) light, the Stokes parameters are trivial
    $$S^{a}_{nat}=(I_{nat} , 0, 0, 0)\; .$$
When summing two non-coherent light waves, their Stokes parameters behave in accordance with the
    following linear law: $I_{(1)}+I_{(2)}\; ,\; {\vec S} _{(1)}+{\vec S}_{(2)} $. In
    particular, partially polarized light can be obtained as linear sum of natural and completely
    polarized light:
    $$S^{a} _{nat}=(I_{nat} , 0, 0, 0)\;, \qquad S^{a}_{pol}=(I_{pol}, I_{pol}\; {\vec n})\;,
    S^{a}=(\; I_{nat}+I_{pol}\;)\; \left (1 , { I_{pol} \over I_{nat}+I_{pol} }\; {\vec n} \right)\;.$$
We further denote
    $$I=I_{nat}+I_{pol}\; , \qquad p={ I_{pol} \over I_{nat}+ I_{pol} }\; ,$$
and then, for the Stokes vector of the partially polarized light we have
    $$S^{a}=(I ,\; I\; p\; {\vec n})\; , \qquad S_{a}S^{a}=I^2 (1-p^{\; 2}) \geq 0\; ,$$
where $I>0$ is the general intensity, $p$ is the degree of polarization (which runs within
    the $[0,\; 1]$ interval: $0 \leq p\leq 1 $), and ${\vec n}$ stands for any unit 3-vector.
    Due to the relations:
    $$\ba{l}S_{a} S^{a}=I^2_{pol}-{\vec S}^{\;2}_{pol}=0 \qquad\mbox{for completely polarized light};\mm
    S_{a}S^{a}=I^2(1-p^2) \geq 0 \qquad\mbox{for partially polarized light},\ea$$
the behavior of Stokes 4-vectors for completely and partially polarized light under acting optic
    devices may be sometimes considered as isomorphic to the behavior of respectively the isotropic
    and the time-like vectors with respect to Lorentz group of Special Relativity:
    $$\ba{l}S_{a} S^{a} =\mbox{inv}= 0 \qquad\mbox{completely polarized light};\mm
    S_{a}S^{a}=\mbox{inv} \geq 0 \qquad\mbox{partially polarized light}\; .\ea$$
This simple observation leads to many consequences, of which some will be discussed below.
%
%
\section{Polarized light and Jones formalism, restriction to the \\$SU(2)$-symmetry,
    and two sorts of non-relativistic Stokes 3-vectors}
Let us consider now the polarization Jones formalism and its connection with spinors for
    rotation and Lorentz groups \cite{1}. It is convenient to start with a relativistic 2-spinor $\Psi$,
    representation of the special linear group $GL(2,\C)$, covering for the Lorentz group
    $L_{+}^{\uparrow}$:
    $$\ba{l}\Psi=\left(\ba{c}\Psi^1\\\Psi^2\ea \right)\; , \qquad\Psi^{\prime}=B(k)\Psi\; ,
        \qquad B(k) \in SL(2,\C)\; ,\mm
    B(k)=k_{0}+k_{j} \sigma^{j}\; , \qquad \mbox{det} B =k_{0}^2- {\bf k}^2=1\; .\ea$$
From the spinor $\Psi$ one may construct a 2-rank spinor $\Psi\otimes\Psi^{*} $, which in turn can
    be resolved in terms of Pauli matrices (we need two sets: $\sigma^{a}=(I, \sigma^{j})$ and
    $\bar{\sigma}^{a}=(I, -\sigma^{j})$):
    $$\Psi\otimes\Psi^{*}={1 \over 2}\;(S_{a}\; \bar{\sigma}^{a})={1 \over 2}\; (S_{0}-S_{j}\; \sigma^{j})\; .$$
The spinor nature of $\Psi$ generates a corresponding (Lorentz) transformation law for $S_{a}$:
    $$S\;^{\prime}_{a}\; \bar{\sigma}^{a}= S_{a}\; B(k) \bar{\sigma}^{a} B^{+}(k)\; ,$$
which - with the use of the well-known relation in the theory of the Lorentz group \cite{1} - can be written:
    $$\ba{l}B(k) \bar{\sigma}^{a} B^{+}(k)=\bar{\sigma}^{b}L_{b}^{\;\;\;\;a} \qquad \Longrightarrow
        \qquad S^{\prime}_{b}=L_{b}^{\;\;\;a}\; S_{a}\; ,\mm
    L^{\;\; a}_{b} (k,\; k^{*})=\bar{\delta }^{c}_{b}\; \left [\;- \delta^{a}_{c}\; k^{n}\; k^{*}_n \;+\;
        k_{c}\; k^{a*}\;+\; k^{*}_{c}\; k^{a}\;+\; i\; \epsilon^{\;\;anm}_{c}\;k_n \; k^{*}_{m}\;
        \right ]\; ,\mm
    \bar{\delta }^{c}_{b}=\left \{\ba{l}+1 ,\;\; c=b=0\; ;\\-1 ,\;\; c=b=1,\; 2,\; 3\; .\ea \right.\ea$$
Thus, the spinor transformation $B(k)$ for the spinor $\Psi$ generates the linear transformation
    $L_{b}^{\;\;\;a} (k, k^{*})$ over Stokes vectors, which preserves the (relativistic) length.
    We note that opposed by sign spinor matrices $\pm B$, lead to the same matrix $L$.\\[2mm]
If we restrict ourselves to the case of the $SU(2)$ group \cite{1}, we get\footnote{We assume here $k_{0}=n_{0}$,
    $\vec{k}=i \vec{n}$.}
    $$L(\pm n)= \left(\ba{rrrr}1&0&0&0\\
    0&1 -2 (n_2^2+n_3^2)&-2n_{0}n_3+2n_1n_2&2n_{0}n_2+ 2n_1n_3\\
    0&2n_{0}n_3+2n_1n_2&1-2(n_1^2 +n_3^2) &-2n_{0}n_1 +2n_2n_3\\
    0&-2n_{0}n_2+2n_1n_3&2n_{0}n_1+2n_2n_3&1 -2(n_1^2+n_2^2)\ea \right)\; .$$
We introduce now a special parametrization for the Jones spinor $\Psi$:
    $$\ba{l}\Psi=\left(\ba{r} N e^{i\alpha}\\ M e^{i\beta}\ea \right) , \qquad\Psi\otimes\Psi^{* }=
    {1 \over 2} \left(\ba{cc}S^{0}+S^3&S^1-i S^2\\ S^1+i S^2 &S^{0}-S^3 \ea \right) ,\mm
    S^1=2NM \cos (\beta-\alpha)\; , \qquad S^2=2NM \sin (\beta-\alpha)\; ,\mm
    S^3=N^2-M^2\; , \qquad S^{0}=N^2+M^2=+ \sqrt { S_1^2 +S_2^2+S_3^2 }\; ;\ea$$
which coincides with the above definition for the case of completely polarized light
    $$\ba{l}S^{0}=N^2+M^2=+ \sqrt { S_1^2 +S_2^2+S_3^2 }\; ,\;\; S^3=N^2-M^2\; ,\mm
    S^1=2NM\;\cos \Delta\; , \qquad S^2=2NM\; \sin \Delta\;.\ea$$
However, there exist two ways to construct a 3-vector in terms of
2-spinors:
    $$\ba{l}(\Psi\otimes\Psi^{*})= r\;+\; x_{j}\; \sigma^{j}\;\; ,\;\;
        r=+ \sqrt{ x_{j}\; x_{j}}\; , \qquad x_{j}-\mbox{pseudovector}\; ;\mm
    (\Psi^{\prime}\otimes\Psi^{\prime})=\;(y_{j}\;+\; i\; x_{j})\;\sigma^{j}\;
        \sigma^2\; , \qquad y_{j},x_{j}-\mbox{vectors}\; .\ea$$
Evidently, the first variant provides us with a possibility to build a spinor model for the
    pseudo-vector 3-space, whereas the second variant leads to a spinor model of a proper vector
    3-space\footnote{According to Cartan, a discret spinor reflection
    is given by the $(2\times 2)$-matrix $i I$.}. Correspondingly, there are possible two Jones
    spinors: $\Psi\; \Longleftrightarrow\; S_{j}\; ,\;\Psi^{\prime}\; \Longleftrightarrow\; S_{j}\; . $
    The Jones-like formulas for Stokes 3-vectors, in both cases, look as follows:
\begin{itemize}\setlength{\parskip}{-1mm}
\item (traditional) $\Psi ({\bf S})$
    $$S^1=\sqrt{{N M \over 2}}\; \cos \Delta\; ,\; S^2=\sqrt{{N M \over 2}}\;
        \sin \Delta\; ,\; S^3=N^2- M^2\; ;$$
\item (alternative) $\Psi^{\prime} ({\bf S})$
    $$\ba{l}S^1=\sqrt{2 \mid M^{^{\prime}2}-N^{^{\prime}2} \mid }\; \cos \Delta\; ,\mm
    S^2=\sqrt{2 \mid M^{^{\prime}2}-N^{^{\prime}2} \mid }\; \sin \Delta\; ,\;
    S^3=\pm\; \sqrt{N^{\prime}M^{\prime}}\; .\ea$$
\end{itemize}
%
%
\section{Spinor representation of Stokes 4-vectors and 2-rank tensors for completely polarized light}
A bi-spinor of second rank $U=\Psi\otimes\Psi $ can be resolved into scalar $\Phi $, a vector
    $\Phi _{b}$, a pseudoscalar $\tilde{\Phi } $, a pseudovector $\tilde{\Phi }_{b} $, and
    a skew-symmetric tensor $\Phi _{ab}$, as follows
    $$\ba{l}U=\Psi\otimes\Psi=\left [-i\; \Phi+\gamma^{b}\;\Phi _{b}+  i\; \sigma^{ab}\;
        \Phi _{ab}+\gamma^{5}\; \tilde{\Phi }+  i\; \gamma^{b} \gamma^{5}\; \tilde {\Phi }_{b}
        \right ] E^{-1}\; ,E= \left(\ba{cc}i \sigma^2&0\\0 &-i \sigma^2 \ea \right) ,\mm
    \gamma^{a}=\left(\ba{cc}0&\bar{\sigma}^{a}\\\sigma^{a}&0 \ea \right)\; ,\;
        \gamma^{5}=\left(\ba{cc}-I&0\\ 0&+I \ea \right)\;,\;\sigma^{ab}={1 \over 4}\; \left(\ba{cc}
        \bar{\sigma}^{a} \sigma^{b}-\bar{\sigma}^{b} \sigma^{a}&0\mm
    0&\sigma^{a} \bar{\sigma}^{b}-\sigma^{b} \bar{\sigma}^{a}\ea \right) .\ea$$
The inverse relations are
    $$\ba{l}\Phi _{a}={1\over 4}\;\mbox{Sp}\; [ E \gamma _{a} U ]\; ,\qquad \tilde{\Phi }_{a}={1\over 4i}\;
        \mbox{Sp}\; [E \gamma^{5}\gamma _{a} U ]\; ,\mm
    \Phi={i \over 4}\;\mbox{Sp}\; [ E U ]\; ,\; \tilde{\Phi }= {1\over 4}\;\mbox{Sp}\;
        [E \gamma^{5} U ]\; ,\;\;\;\Phi_{mn}=-{1 \over 2i}\;\mbox{Sp}\; [E \sigma _{mn} U ]\; .\ea$$
The explicit expressions for tensors obtained from spinors are
    $$\ba{l}\Phi_{0}=\xi^1 \eta_{\dot{2}}-\xi^2 \eta_{\dot{1}}\;,\qquad \Phi_1=
        \xi^1 \eta_{\dot{1}}-\xi^2\eta_{\dot{2}}\; ,\mm
    \Phi_2= i\; (\xi^1 \eta_{\dot{1}}+\xi^2 \eta_{\dot{2}})\; ,\qquad
        \Phi_3=-\; (\xi^1 \eta_{\dot{2}}+\xi^2\eta_{\dot{1}})\; ,\mm
    \tilde{\Phi}_{0}=0\; ,\; \tilde{\Phi}_1=0\; ,\;
        \tilde{\Phi}_2=0\;,\; \tilde{\Phi}_3=0\; ,\; \Phi=0\;,\; \tilde{\Phi}=0\; ,\ea$$
and
    $$\ba{l}\Phi^{01}={i \over 4}\; [\; (\xi^1\xi^1-\xi^2\xi^2)+
        (\eta_{\dot{1}} \eta_{\dot{1}}-\eta_{\dot{2}}\eta_{\dot{2}})\; ]\; ,\mm
    \Phi^{23}={1 \over 4}\; [\; (\xi^1\xi^1-\xi^2\xi^2)-(\eta_{\dot{1}} \eta_{\dot{1}}-\eta_{\dot{2}}
        \eta_{\dot{2}})\; ]\; ,\mm
    \Phi^{02}=- {1 \over 4}\; [\; (\xi^1\xi^1+\xi^2\xi^2)+(\eta_{\dot{1}} \eta_{\dot{1}}+\eta_{\dot{2}}
        \eta_{\dot{2}})\; ]\; ,\mm
    \Phi^{31}=- {1 \over 4i}\; [\; (\xi^1\xi^1+\xi^2\xi^2)-(\eta_{\dot{1}} \eta_{\dot{1}}+
        \eta_{\dot{2}} \eta_{\dot{2}})\; ]\; ,\mm
    \Phi^{03}=- {i \over 2 }\; [\;\xi^1\xi^2+\eta_{\dot{1}} \eta_{\dot{2}} ]\; ,\qquad \Phi^{12}=
        -{1\over 2 }\; [\;\xi^1\xi^2-\eta_{\dot{1}}\eta_{\dot{2}} ]\; .\ea$$
By collecting the results, we infer:
    $$\Psi=\left(\ba{c}\xi^{\alpha}\\\eta_{\dot{\alpha}}\ea \right)\; ,\Psi\otimes\Psi
        \Longrightarrow\Phi =0 ,\; \tilde{\Phi}=0,\; \tilde{\Phi}_{a}=0 ,\;
        {\Phi}_{a} \neq 0 ,\; {\Phi}_{mn} \neq 0\; .$$
In order to obtain the vector and the tensor both real, one should impose additional restrictions:
    $$\eta=- i\; \sigma^2\;\xi^{*}\; \qquad \Longrightarrow\qquad \eta_{\dot{1}}=
        -\xi^{2*}\; ,\;\; \eta_{\dot{2}}=+\xi^{1*}\; ,$$
which results in
    $$\ba{l}\Phi_{0}=(\xi^1\;\xi^{1*}+\xi^2\;\xi^{2*})> 0\; , \qquad \Phi_3=
        - (\xi^1\;\xi^{1*}-\xi^2\;\xi^{2*})\; ,\mm
    \Phi_1 =- (\xi^1\;\xi^{2*}+\xi^2\;\xi^{1*})\; , \qquad
        \Phi_2=-i\; (\xi^1\;\xi^{2*}-\xi^2\;\xi^{1*})\; ;\mm
    \Phi^{01}={i \over 4}\; [\; (\xi^1\;\xi^1-\xi^2
        \;\xi^2)+(\xi^{2*}\;\xi^{2*}-\xi^{1*}\;\xi^{1*})\;\; ,\;\;\mbox{and so on}.\ea$$
The last case seems to be the most appropriate to describe Stokes 4-vectors and to
    determine the Stokes 2-rank tensor. The main invariant turns to equal to zero, since:
    $$S_{0}S_{0}-S_{j}S_{j}=0\; ,$$
and hence $S_{a}$ may be considered as a Stokes 4-vector for completely polarized light.\pas
In turn, the 4-tensor $S_{mn}$, being constructed from Jones bi-spinor $\Psi$, is a Stokes
    2-rank tensor. We further calculate the two invariants for $S_{mn}$:
    \zx{inv-s}I_1=-{1 \over 2}\; S^{mn}S_{mn}=0\; , \qquad I_2={1\over 4}\; \epsilon_{abmn} S^{ab} S^{mn}=0\;.\zc
Instead of the Stokes 4-tensor $S_{ab}$, one may introduce a complex 3-vector,
    $$\ba{l}s^1=S^{01}+i S^{23}\; ,\; s^2=S^{02}+i S^{31}\;,\; s^3=S^{03}=i S^{12}\;,\mm
    s_1+i s_2=- i\;\xi^2\xi^2\; , \qquad s_1-is_2=+ i\;\xi^1\xi^1\; , \qquad s^3-i\;\;\xi^1\;\xi^2\; .\ea$$
Additionally to Jones spinor and Mueller vector formalisms, the later considerations allow
    to introduce one other technique, which is based on the use of complex 3-vectors, under the
    complex rotation group $SO(3,\C)$: This complex vector is isotropic, ${\bf s}^2=0 $.
%
%
\section{The Jones 4-spinor for partially polarized light}
Now let us examine one more possibility of combining two spinors:
    $$\Psi\otimes (-i\Psi^{c})=\left(\ba{c}\xi^1\\\xi^2\\\eta_{\dot{1}}\\ \eta_{\dot{2}}\ea \right)
        \otimes \left(\ba{c}+ \eta_{\dot{2}}^{*}\\-\eta_{\dot{1}}^{*}\\-\xi^{2*}\\+\xi^{1*}\ea \right).$$
With the notation
    $$\xi=\left(\ba{c}N_1 e^{in_1}\\N_2 e^{in_2} \ea \right)\; , \qquad \eta=\left(\ba{c}
        M_1 e^{im_1}\\M_2 e^{im_2} \ea \right)\; ,$$
we can prove that the corresponding 4-vector is time-like:
    $$(N_1 M_1-N_2 M_2)^2<\Phi_{0}^2-\vec{\Phi}^2<(N_1 M_1+N_2 M_2)^2\; .$$
This means that we have ground to consider the 4-vector $\Phi_{a}$ as a Stokes 4-vector $S_{a}$.
    Therefore, the 4-spinor is of Jones type and corresponds to partially polarized light.\pas
It remains to explicitly find the form for the corresponding (real) Stokes 4-tensor $S_{ab}$;
    its description with the help of complex 3-vectors looks most simple:
    $$s^1={i \over 2} (\xi^1\eta_{\dot{2}}^{*}+\xi^2\eta_{\dot{1}}^{*})\; ,\;
        s^2=- {1 \over 2}(\xi^1\eta_{\dot{2}}^{*}-\xi^2 \eta_{\dot{1}}^{*})\; ,\;
        s^3=- {i \over 2} (\xi^2 \eta_{\dot{2}}^{*} -\xi^1\eta_{\dot{1}}^{*})\; ;$$
this complex 3-vector is not isotropic,
    $${\bf s}^2=-{1 \over 4}\; (\xi^1 \eta_1^{*}-\xi^2\eta_1^{*})^2 \neq 0\; .$$
One more last remark should be added: the results of Sections 1--4 can be of use not only in
    polarization optics, but also they may  be of interest to describe Maxwell theory
    in spinor approach, when instead of variables $A_n , F_{mn}$ one introduces one fundamental
    electromagnetic bi-spinor $\Psi=(\xi , \eta)$. As well, these results can have a meaning
    in the context of explicitly constructing relativistic models for space-time with spinor structure.
%
%
\section{The linear group $SL(4,\R)$ and the classification of 1-parametric Mueller matrices}
The main goal of this section is to develop a systematic method of identifying and classifying
    the Mueller matrices within the family of matrices of the real group $SL(4,\R)$.
We note that to construct the general transformation of the group $SL(4,\R)$ is straightforward,
    but to analyze the adequacy of such a transformation for describing Mueller matrices is a
    highly nontrivial (practically impossible) task. However, using the technique of Dirac matrices,
    we can, quite easily explicitly describe all the 16 one-parametric subgroups, from which, using
    all the possible emerging products, one can produce the whole group $SL(4,\R)$.
    For these distinct 1-parametric subgroups, the question of their adequacy of being Mueller
    matrices becomes sufficiently simple, and thus we obtain in each case a definite answer.
    In particular, diagonal subgroups are trivially simple and will not be further discussed as
    subcase of valid Mueller solutions.
Any Mueller matrix of general type, $M _{ab} S_{a}=S_{a}^{\prime}$, must obey the following restrictions
    $$\ba{l}S_{0} \geq 0\;, \qquad S^2 \equiv S^2_{0}-S^2_1-S^2_2-S^2_3 \geq 0\; ,\mm
        S^{\prime}_{0} \geq 0\;, \qquad S^{\prime}{}^2 \equiv S^{\prime}{}^2_{0}-
            S^{\prime}{}^2_1- S^{\prime}{}^2_2-S^{\prime}{}^2_3 \geq 0\; ,\ea$$
or, in more detailed form,
    $$\ba{l}M_{00} S_{0}+M_{01} S_1+M_{02} S_2+M_{03} S_3 \geq 0\; ,\mm
    (M_{00} S_{0}+M_{01} S_1+M_{02} S_2+M_{03} S_3)^2\mm
    -(M_{10} S_{0}+M_{11} S_1+M_{12} S_2+M_{13} S_3)^2\mm
    - (M_{20} S_{0}+M_{21} S_1+M_{22} S_2+M_{23} S_3)^2\mm
    -(M_{30} S_{0}+M_{31} S_1+M_{32} S_2+M_{33} S_3)^2\geq 0\; .\ea$$
We shall further use the following notation:
    $$S_{0} =I, S_{j}=I p_{j} ,\; p_1=a ,\; p_2=b ,\;p_3=c\; .$$
For describing the change of the degree of polarization, one can use the quantity $D$:
    $$(a^{^{\prime}2}+b^{^{\prime}2}+c^{^{\prime}2})-(a^2+b^2+c^2)=D.$$
No we are ready to specify the 12 non-diagonal 1-parametric subgroups in $SL(4,\R)$.\pam
\vv{Variant (1):}
    $$M= U_1^{\alpha} (\phi)=\left(\ba{cccc}\cos \phi&\sin \phi&0&0\\-\sin \phi&\cos \phi&0&0\\
    0&0&\cos \phi&-\sin \phi\\ 0&0&\sin \phi&\cos \phi\ea \right) ,$$
where the restrictions (in the variables $ \tan \phi=x $) look like
    $$ a \sin \phi+\cos \phi \geq 0\; ,\;\;{1-x^2 \over 1+x^2 } (1-a^2)+{2 x
        \over 1+x^2 }\; 2 a-b^2-c^2 \geq 0\; ,$$
and where the solution depends on the initial Stokes vector and is much simplified
    in the case of completely polarized light: $x \in [ x_1, x_2]$, where
    $$\ba{l}x_1={ 2a-\sqrt{ 4a^2+(1 -p^2)\; (b^2+c^2+1-a^2) } \over b^2+c^2+1-a^2}\; ,\mm
        x_2={ 2a+\sqrt{ 4a^2+(1 -p^2)\; (b^2+c^2+1-a^2) } \over b^2+c^2+1-a^2}\; .\ea$$
The possible values of the parameter $D$ lead to subcases:
    $$\ba{l}D<0 , \qquad \Longrightarrow \qquad 0<\tan \phi<{2a \over 1-a^2}\;\mbox{(decreasing)}\;,\mm
    D>0 , \qquad \Longrightarrow \qquad \tan \phi > {2a \over 1-a^2}\;\mbox{(increasing)}\; ,\mm
    D=0 \qquad \Longrightarrow \qquad \tan \phi= {2a \over 1 -a^2}\;\mbox{(non-changing)}\; .\ea$$
We note that this result is typical for all six one-parametric
subgroups ({\bf Variants 1--6}) in the following sense:
    the appropriateness of the elementary matrix $M$ to be of Mueller type depends on the
    parameters of the matrix and on the characteristics of the initial light beam.
    Hence, when combining more complex Mueller matrices by multiplying elementary 1-parametric
    ones, we must check each next step of the chain
    $$(\ldots M_n  M_{n-1}\ldots M_2 M_1)\; S=S^{\prime}\; .$$
\vv{Variant (2):}
    $$M=U_2^{\alpha} (-\phi)=\left(\ba{cccc}\cos \phi&0&\sin \phi&0\\0&\cos \phi&0&\sin \phi\\
        -\sin \phi&0&\cos \phi&0\\0&-\sin \phi&0&\cos \phi\ea \right) .$$
The restrictions are the following
    $$\cos \phi+b\; \sin \phi \geq 0\; ,\;\;\; {1-x^2 \over 1+x^2 } (1 -b^2)+{2 x \over
        1+x^2 }\; 2 b-a^2-c^2 \geq 0 ,$$
and they differ from the previous ones only by notation.\\[3mm]
\vv{Variant (3):}
    $$\ba{l}M=U_3^{\alpha} (\phi)=\left(\ba{cccc}\cos \phi&0&0&\sin \phi\\
        0&\cos \phi &-\sin \phi&0\\0&\sin \phi&\cos \phi&0\\- \sin \phi&0&0&\cos \phi\ea \right) ,\mm
    \cos \phi+c\; \sin \phi \geq\; , \qquad{1-x^2 \over 1+x^2 } (1-c^2)+{2x \over 1+x^2 }
        \; 2 c-a^2-b^2 \geq 0\; .\ea$$
\vv{Variant (4):}
    $$\ba{l}M =U_1^{\beta}(\phi)=\left(\ba{cccc}\cos \phi&\sin \phi&0&0\\-\sin \phi&\cos \phi&0&0\\
        0&0&\cos \phi&\sin \phi\\ 0&0&-\sin \phi&\cos \phi\ea \right) ,\mm
    \cos \phi+ a\; \sin \phi \geq 0\; , \qquad{1-x^2 \over 1+x^2 }(1-a^2)+{2 x \over 1+x^2}
        \; 2 a-b^2-c^2 \geq 0\; .\ea$$
\vv{Variant (5):}
    $$\ba{l}M=U_2^{\beta} (\phi)=\left(\ba{cccc}\cos \phi&0&\sin \phi&0\\0&\cos \phi&0&-\sin \phi\\
        -\sin \phi&0&\cos \phi&0\\0&\sin \phi&0&\cos \phi\ea \right) ,\mm
    \cos \phi-b\; \sin \phi \geq 0\; ,\qquad {1-x^2 \over1+x^2 } (1 -b^2)+{2 x \over 1+x^2 }\;
        2 b-a^2-c^2 \geq 0\; .\ea$$
\vv{Variant (6):}
    $$\ba{l}M=U_3^{\beta} (\phi)=\left(\ba{cccc}\cos \phi&0&0&\sin \phi\\0&\cos \phi&\sin \phi&0\\
        0&-\sin \phi&\cos \phi&0\\- \sin \phi&0&0&\cos \phi\ea \right) ,\mm
    \cos \phi+c\; \sin \phi \geq 0\; , \qquad {1-x^2 \over1+x^2 } (1 -c^2)+{2 x \over 1+x^2 }\;
        2 c-a^2-b^2 \geq 0\; .\ea$$
Next, we will consider six  one-parametric subgroups  constructed
with the use of hyperbolic functions.

\vv{Variant (7):}
    $$U_2^{A} (-i\beta)= \left(\ba{cccc}\cosh\;\beta&0&0 &\sinh\;\beta\\
        0 &\cosh\;\beta&-\sinh\;\beta&0\\0&-\sinh\;\beta &\cosh\;\beta&0\\
        \sinh\;\beta&0&0 &\cosh\;\beta\ea \right),$$
for which we note that the restriction $\cosh\;\beta S_{0}+\sinh\;\beta S_3 \geq 0$ is valid
    for arbitrary $\beta$.\pas
The quadratic inequality in the variables $a,b,c$ and $y=\mbox{th}\;\beta, y \in (-1, +1)$,
    takes the form
    $$-y^2 (a^2+b^2 +1 -c^2)+4ab y+(1 -a^2-b^2 -c^2)\geq 0\; ,$$
with the solution
    $$\ba{l}y \in [ y_1 , y_2 ]\;,\mm
    y_1={2ab-\sqrt{4a^2b^2+(1 -p^2)(a^2+b^2 +1-c^2)} \over a^2+b^2 +1-c^2}<0\; ,\mm
    y_2={2ab+\sqrt{4a^2b^2+(1 -p^2)(a^2+b^2 +1-c^2)} \over a^2+b^2 +1-c^2} > 0\; .\ea$$
The results depend on the initial light. For completely polarized light, the formulas become much
    simpler. The degree of polarization changes according to the rules
    $$D={ (a -b y)^2+(b -ay)^2+(c+y)^2 \over (1+c y)^2}- a^2-b^2-c^2\; .$$
This result is typical again for these six cases in the sense described above.\\[3mm]
\vv{Variant (8):}
    $$U_3^{A} (i\beta)= \left(\ba{cccc}\cosh\;\beta&0&-\sinh\;\beta&0\\
    0 &\cosh\;\beta&0&-\sinh\;\beta\\-\sinh\;\beta&0 &\cosh\;\beta&0\\
    0&-\sinh\;\beta&0 &\cosh\;\beta\ea \right).;$$
\vv{Variant (9):}
    $$U_1^{B} (i\beta)= \left(\ba{cccc}\cosh\;\beta&0&0&-\sinh\;\beta\\
    0 &\cosh\;\beta&-\sinh\;\beta&0\\0&-\sinh\;\beta &\cosh\;\beta&0\\
    -\sinh\;\beta&0&0 &\cosh\;\beta\ea \right) .$$
\vv{Variant (10):}
    $$U_3^{B} (i\beta)=\left(\ba{cccc}\cosh\;\beta &\sinh\;\beta&0&0\\
    \sinh\;\beta &\cosh\;\beta&0&0\\0&0 &\cosh\;\beta &-\sinh\;\beta\\
    0&0&-\sinh\;\beta &\cosh\;\beta\ea \right) .$$
\vv{Variant (11):}
    $$U_1^{C} (i\beta)= \left(\ba{cccc}\cosh\;\beta&0 &\sinh\;\beta&0\\
    0 &\cosh\;\beta&0 &-\sinh\;\beta\\\sinh\;\beta&0 &\cosh\;\beta&0\\
    0 &-\sinh\;\beta&0 &\cosh\;\beta\ea \right).$$
\vv{Variant (12):}
    $$U_2^{C} (-i\beta)=\left(\ba{cccc}\cosh\;\beta&-\sinh\;\beta&0&0\\
    -\sinh\;\beta &\cosh\;\beta&0&0\\0&0 &\cosh\;\beta&-\sinh\;\beta\\
    0&0&-\sinh\;\beta &\cosh\;\beta\ea \right).$$
The appropriateness of the elementary matrix $M$ to be of Mueller type depends on the
    parameters of the matrix and on the characteristics of the initial light beam.
    While producing a more complex Mueller matrix by multiplying elementary 1-parametric
    Mueller matrices, we must check each next step in the chain
    $$(\ldots M_n  M_{n-1}\ldots M_2 M_1)\; S=S^{\prime}\;.$$
%
%

\section{The semi-group structure and classification of degenerate Mueller matrices;
    projective geometry}
\vv{Preliminary remarks}. The Mueller transformation formulas
    $p_{j} \Longrightarrow p_{j}^{\prime}$ can be presented as a law of a
    projective (15-parametric) group:
    $$\ba{l}p^{\prime}_1={m_{10}+m_{11} p_1+m_{12} p_2+m_{13} p_3\over 1+m_{01} p_1+m_{02}
        p_2+m_{03} p_3}\;,\mm
    p^{\prime}_2={m_{20}+m_{21} p_1+m_{22} p_2+m_{23} p_3\over 1+m_{01} p_1+m_{02} p_2+
        m_{03} p_3}\;,\mm
    p^{\prime}_3={m_{30}+m_{31} p_1+m_{32} p_2+m_{33} p_3\over 1+m_{01} p_1+m_{02} p_2+
        m_{03} p_3}\;;\ea $$
with the constraints
    $$\ba{l}1+m_{01} p_1+m_{02} p_2+m_{03} p_3>0\;,\mm
    p_1^2+p_2^2+p_3^2 \leq 1\; ,\qquad p_1^{^{\prime}2}+p_2^{^{\prime}2}+p_3^{^{\prime}2} \leq 1\;.\ea$$
With respect to a spinor basis, any $4\times4$ matrix can be constructed by means of four
    4-dimensional objects (vectors) $(k,m,l,n)$, as follows
    $$\left(\ba{cc}k_0+\; {\bf k}\; \vec{\sigma}&n_0+\; {\bf n}\; \vec{\sigma}\\[3mm]
        \ell_0+\; {\bf l}\; \vec{\sigma}&m_0+\; {\bf m}\; \vec{\sigma}\ea\right)=
        \left(\ba{cc}K&N\\[3mm] L&M \ea\right); $$
where we use the notation $k=(k_{0},k_{j})$ and so on. The symbol $\vec{\sigma}= (\sigma_{j})$
    stands for the three $ 2\times 2 $ Pauli matrices. The four $2 \times 2$ blocks are denoted as $K,M,L,N$.\pas
In order to have matrices with real elements, it is necessary to require that the components which have
    the index 2, to be imaginary:
    $$ k_2 -> ik_2\;,\qquad m_2 -> im_2\;,\qquad n_2 ->in_2\;,\qquad l_2 -> il_2\; , $$
leaving real the other components of the parameters.\pas
By imposing linear constraints on the four 4-dimensional vectors, and by requiring that the
 group law for multiplication is valid for these parameters \cite{9, 9'},
 we can obtain a large variety of simple subsets of matrices \cite{8, 6}. All of them have a definite
 mathematical structure: either of sub-group or of semi-group. A large part of these subsets
 consist of degenerate matrices. Otherwise speaking, one might obtain in this manner a large
 number of semigroups of 4-th order matrices (more than 40 -- see \cite{8, 6}). However, the question of adequacy
 of such simple subsets of matrices for describing Mueller transformations has not been
 addressed until now. The purpose of this section is to perform such an analysis.\pas
Below we shall present only a few typical examples of these sets.\pas
\vv{One single independent vector $(k_0,{\bf k})$}. We shall examine the case when
    the independent 4-dimensional vector is $(k_0, {\bf k})$:
    $${\bf n}=A\; {\bf k}\;,\; n_0=\alpha\; k_0\;,\;
     {\bf m}=B\; {\bf k}\;,\; m_0=\beta\; k_0\;,\;
     {\bf l}=D\; {\bf k}\;,\; l_0=t\; k_0\;.$$
In this case, by imposing the requirement of satisfying the axioms of group law provides
    7 distinct solutions: K1--K7 (see \cite{8, 6}), as described below.\pamn
\vv{Variant K1:}
    $$G=\left(\ba{cc} K&0\\0&0 \ea\right)=\left(\ba{cccc}k_0+k_3&k_1+k_2&0&0\\k_1-k_2&k_0-k_3&0&0\\
    0&0&0&0\\0&0&0&0\ea\right)=\left(\ba{cccc}a&c&0&0\\d&b&0&0\\0&0&0&0\\0&0&0&0\ea\right);$$
where all the $(4\times 4)$-matrices are degenerate. The rank of such a matrix is either 2 or 1
    (while in the last case one should require $\det\; K= ab -cd=0$).
    Transformations are of Mueller type only if
    $$ a+c x>0\;,\qquad x \in [-1, 1]\;, \qquad(cx+a)^2- (bx+d)^2 \geq 0\;.$$
The projective transformation has the form
    $$x^{\prime}={d+bx \over a+c x},\;\; y^{\prime}=y,\;\; z^{\prime}=z\; ,$$
which leads to two systems of inequations
    $$\ba{l}I \qquad a+c x>0\;, \qquad (c-b)x+a-d\geq 0\; , \qquad(c+b)x+a+d \geq 0\;;\mm
    II \qquad a+c x>0\;, \qquad (c-b)x+a-d \leq 0\; , \qquad (c+b)x+a+d \leq 0\;,\ea$$
where system II has no solutions.\pas
An important point concerns the appropriateness of these matrices to be of Mueller type.
    This depends on the properties of the initial light beam. The roots of the above quadratic equation are
    $$x_{1,2}={ (ac-bd) \mp (ab-cd)\over b^2- c^2 }\;.$$
If the coefficient $(c^2-b^2)$ at $x^2$ is negative, then the solution of the inequation has the form
    $$x\in[x_1, x_2]\;.$$
If this coefficient is positive, then the solution is of the form
    $$x\in (-\infty; x_1] \cup [x_2,+\infty)\;.$$
It makes sense to impose the requirement $ \det K=ab -c d =+1 $\footnote{This happens due to
    the fact that the norming by the determinant can be always considered, by using a factor
    applied to the matrix $K$}. Then the formulas for the roots simplify to
    $$ x_{1,2}={ \mp 1+(ac-bd)\over b^2- c^2 }\;.$$
Moreover, we can separately tract the case of matrices of rank 1; to this
    aim we need to impose the condition
    $$ab-cd=0 \qquad \Longrightarrow \qquad d={ab\over c}\;,$$
which leads to a very special projective transformation
    $$ x^{\prime}={d+bx\over a+c x}={ ab/c+b x\over a+cx }={b\over c}=
        {d\over a}=\mu\;,\qquad \mid \mu \mid \leq 1\;. $$
For this case, the requirements for being Mueler type matrices are
    $$ a+c x>0\;, \qquad 1 -{b\over c }\geq 0\;, \qquad 1+{b\over c }\geq 0\;,$$
where the last two inequalities are equivalent to $\mid \mu \mid \leq 1$.\\[3mm]
\vv{Variant K2:}
    $$G=\left(\ba{cccc}k_0+k_3&k_1+k_2&0&0\\k_1-k_2&k_0-k_3&0&0\\0&0&k_0+k_3&k_1+k_2\\
    0&0&k_1-k_2&k_0-k_3\ea\right)=\left(\ba{cccc}a&c&0&0\\d&b&0&0\\0&0&a&c\\0&0&d&b\ea\right).$$
This set consists of non-degenerate matrices. By imposing the conditions $\det K=0$, we get a
    semi-group of rank 1. The corresponding projective transformation is given by:
    $$x^{\prime}={d+b x\over a+c x}\;,\qquad y^{\prime}={ay+c z\over a+cx}\; ,\qquad
        z^{\prime}={ay+bz\over a+c x}\; .$$
While limiting ourselves to degenerate matrices of rank 1 ($ab-cd=0$), we get a simpler projective transformation
    $$x^{\prime}={b\over c}\;,\qquad y^{\prime}={ay+c z\over a+cx}\;, \qquad
        z^{\prime}={b\over c}\; {ay+cz\over a+c x}= {b\over c}\; y^{\prime}\;.$$
For Mueller transformations, the following conditions should be fulfilled
    $$\aru{a+c x>0\;,\qquad x^2 +y^2+z^2 \leq 1\;,\mm
        (c^2-b^2) x^2 +2 (ac -bd) x- (a^2+d^2)y^2-(c^2+b^2) z^2 -2 (ac+bd)yz +a^2- d^2\geq 0\;.}$$
We notice that the obtained quadratic inequalities can be considerably simplified
 if we limit ourselves to matrices of rank 1:
    $$\aru{a+c x>0\;,\qquad x^2+y^2+z^2 \leq 1\;,\mm
    \left(1 -{b^2\over c^2}\right) (cx+a)^2-\left(1+{b^2\over c^2}\right)(a y+cz)^2\geq 0\; .}$$
We must assume that $b^2<c^2$, and consequently we get
    $$\ba{l}\sqrt{1 -{b^2\over c^2} } (cx+a)-\sqrt{1+{b^2\over c^2} } (cz+a y) \geq 0\;,\mm
    \sqrt{1 -{b^2\over c^2} } (cx+a)+\sqrt{1+{b^2\over c^2} } (cz+a y)\geq 0\;.\ea$$
We shall examine several more such special particular cases:
    $$\ba{l}x=+1\;,\; y=0\;,\; z= 0\;,\qquad a+c>0\;,\;(a+c)^2\geq (b+d)^2\;;\mm
    x=-1\;,\; y=0\;,\; z= 0\;,\quad a-c>0\;,\;(a-c)^2\geq (b-d)^2\;;\mm
    x=0\;,\; y=+1\;,\; z= 0\;,\qquad a>0\;,\; d=0\;;\mm
    x=0\;,\; y=-1\;,\; z= 0\;, \qquad a>0\;,\; d=0\;;\mm
    x=0\;,\; y=0\;,\; z=+1\;,\qquad a>0\;,\; a^2\geq b^2+c^2+d^2\;;\mm
    x=0\;,\; y=0\;,\; z= -1\;,\qquad a>0\;,\; a^2\geq b^2+c^2+d^2\;\;.\ea$$
In the general case we get the quadratic inequality
    $$\ba{l}x^2+y^2+z^2 \leq 1\;,\qquad a+c x>0\;,\mm
    (a+c x)^2-(d+bx)^2 -(a y+c z)^2- (d y+bz)^2 \geq 0\;.\ea$$
This quadratic form can be diagonalized (we omit the details of this procedure).
    Let us express the fundamental constraint $x^2+y^2+z^2 \leq 1\;$ in terms of new variables $X,Y,Z$.
    We get
    $$\left(X-{ac-bd\over c^2-b^2}\right)^2+Y^2+Z^2 \leq 1\; .$$
The linear inequality $a+cx> 0$ gets the form
    $$c X- { b\; \det K\over c^2-b^2}>0\;.$$
We see, that the task of description of all Mueller matrices of this type is
    solvable, and it is a quite definite problem in the frames of a particular
    projective group.\\[3mm]
\vv{Variant K3}
    $$G=\left(\ba{cc}K&0\\DK&0\ea\right),\qquad G^{\prime}G=
        \left(\ba{cc}K^{\prime}K&0\\DK^{\prime}K&0\ea\right);$$
here $D$ is an arbitrary numeric parameter. This set of matrices is a set of degenerate
    matrices of rank 2 with the structure of a semi-group. We start with
    $$G=\left(\ba{cccc}a&c&0&0\\d&b&0&0\\D a&Dc&0&0\\D d&D b&0&0\ea\right);$$
then the corresponding projective transformation looks like:
    $$x^{\prime}={d+b x\over a+c x}\;,\qquad y^{\prime}=D\;,\qquad z^{\prime}=
        D {d+b x\over a+c x}= D x^{\prime}\;.$$
By limiting ourselves to the semi-group of rank 1, the projective transformation becomes simpler:
    $$\det K=0\;,\qquad x^{\prime}={b\over c }={d \over a}\;,\qquad
        y^{\prime}=D\;,\qquad z^{\prime}=D x^{\prime}\;.$$
The restrictions for having Mueler matrices are
    $$\ba{l}a+c x>0\;,\qquad A x^2+2Bx+C\geq 0\;,\mm
    A= (1-D^2)c^2 -(1+D^2) b^2\;,\mm
    B= (1-D^2) ac-(1+D^2) bd\;,\mm
    C=(1 -D^2) a^2- (1+D^2) d^2\;.\ea$$
The roots of this quadratic equation are
    $$ x_{1,2}= { -bd(1+D^2)+ac(1-D^2) \pm\sqrt{(ab-cd)^2(1-D^{4})}
        \over b^2(1+D^2)-c^2(1-D^2) }\;.$$
If $A > 0 $ (positive) then $x\in[x_1, x_2]$, and if $A<0$, then
    $x\in (-\infty; x_1]\; \cup\;\; [x_2,+\infty)$. The requirement of having real
    roots $x_{1, 2}$ leads to $ D^2 \leq 1$. In particular, if $D^2=1$, the inequalities
    from above get the form
    $$\aru{-(1+1) (bx+d)^2\geq 0\qquad \Longrightarrow\qquad x=- {d\over b}\;.\mm
    a+cx>0\qquad \Longrightarrow \qquad a -{d \over b} c > 0\; .}$$
In the case of zero determinant $\det K=0$, we get
    $$a+cx>0\;,\qquad \left [(1-D^2) -(1+D^2) {b^2\over c^2}\right ] (a+cx)^2\geq 0\;.$$
We note that the Mueller matrix identifying task involves many details, which are physically interpretable
    within polarization optics, and at the same time are relevant in terms of properties of special
    projective transformations. There exist yet about 40 special cases of matrices (mainly with
    semi-group structure -- see \cite{8, 6}) which provide special projective transformations and can describe sets
    of Mueller matrices.
%
%
\section*{Acknowledgment}
The present work was developed under the auspices of Grant 1196/2012 - BRFFR-RA No.
    F12RA-002, within the cooperation framework between Romanian Academy
    and Belarusian Republican Foundation for Fundamental Research.\par
The authors wish to thank to the organizers of the joint event {\em Colloquium on Differential Geometry},
    and {\em The IX-th International Conference on Finsler Extensions of
    Relativity Theory (FERT 2013)}, held between 26 -- 30 August 2013 in Debrecen, Hungary,
    for their worm hospitality.
Also, V. Red'kov, O. Veko and V. Balan are thankful to Prof. D. Pavlov for the support
    provided for the participation in the event {\em FERT 2013}.
\noindent
Elena Ovsiyuk, Olga Veko\\Mozyr State Pedagogical University, Belarus.\\[1mm]
Mircea Neagu\\University Transilvania of Bra\c{s}ov, Romania.\\[1mm]
Vladimir Balan\\University Politehnica of Bucharest, Romania.\\[1mm]
Victor Red'kov\\B.I. Stepanov Institute of Physics, NAS of Belarus.
\end{document}